\documentclass{Interspeech}

\interspeechcameraready

\usepackage{color, colortbl, arydshln}
\usepackage{pifont}
\title{Bringing Interpretability to Neural Audio Codecs}

\author[affiliation={1}, equalcontribution]{Samir}{Sadok}
\author[affiliation={2,3}, equalcontribution]{Julien}{Hauret}
\author[affiliation={2}]{Éric}{Bavu}

\affiliation{Inria}{Université Grenoble Alpes CNRS, LJK}{France}
\affiliation{LMSSC}{Conservatoire national des arts et métiers}{Paris, France}
\affiliation{APC}{French-German Research Institute of Saint-Louis}{France}
\email{
{\normalfont\normalsize\textsuperscript{*}Equal contribution}\\
samir.sadok@inria.fr, julien.hauret@lecnam.net, eric.bavu@lecnam.net
}

\usepackage{subcaption}

\begin{document}

\maketitle

\begin{abstract}

The advent of neural audio codecs has increased in popularity due to their potential for efficiently modeling audio with transformers. Such advanced codecs represent audio from a highly continuous waveform to low-sampled discrete units. In contrast to semantic units, acoustic units may lack interpretability because their training objectives primarily focus on reconstruction performance. This paper proposes a two-step approach to explore the encoding of speech information within the codec tokens. The primary goal of the analysis stage is to gain deeper insight into how speech attributes such as content, identity, and pitch are encoded. The synthesis stage then trains an AnCoGen network for post-hoc explanation of codecs to extract speech attributes from the respective tokens directly.
\end{abstract}

\section{Introduction}
\label{sec:intro}
In the era of deep learning, audio representations are usually built using self-supervised learning. However, there is a notable distinction in whether the task is generating or understanding audio. In speech understanding, notable works such as HuBERT \cite{hsu2021hubert} and WavLM \cite{chen2022wavlm} employ a strategy of leveraging a transformer network's partially masked output for high-level speech comprehension and semantic token generation. Another category of representations has emerged with Soundstream \cite{zeghidour2021soundstream} and Encodec \cite{defossez2022high}, with acoustic tokens produced by neural codecs primarily created for audio compression. These acoustic tokens have since become the de facto building blocks of generation tasks, as they retain all the necessary acoustic details for constructing high-quality audio, unlike semantic tokens. The present paper proposes a study of four contemporary codecs capable of compressing and decompressing audio while maintaining a high level of quality. Each of these codecs possesses a distinctive feature :
\begin{itemize}
\item DAC \cite{kumar2024high}: The first codec to exceed Encodec and Soundsteam in terms of quality, facilitated by two pivotal techniques: factorized codes and L2-normalized codes, which enhance codebook usage. The model is trained with adversarial loss and feature matching on discriminator embeddings while learning codebooks like the original VQ-VAE formulation \cite{van2017neural}. Additionally, the classic residual vector quantization (RVQ) is employed to derive a two-dimensional discrete representation of the audio data.
\item SpeechTokenizer \cite{zhang2023speechtokenizer} enhances the DAC framework by introducing a hierarchical disentanglement of speech information within its RVQ structure. A semantic teacher, such as HuBERT, enables the first RVQ layer to focus on phonetic content, while subsequent layers capture paralinguistic features such as timbre and prosody. This design ensures strong phonetic discriminability, as reflected in a superior Phone-Normalized Mutual Information compared to systems such as Encodec.
\item Mimi \cite{defossez2024moshi} also employs distillation techniques, utilising a split RVQ to circumvent semantic leakage at higher RVQ scales. Additionally, it integrates a transformer network into both the encoder and decoder, while maintaining full causality. It has been demonstrated to outperform both SpeechTokenizer and DAC regarding quality and bitrate.
\item Bigcodec \cite{xin2024bigcodec} is distinct from previous codecs in that it employs a single VQ with a higher cardinality and sampling rate, yielding a similar bitrate as Mimi. The quality of Bigcodec is aligned with the standards of Mimi, with a level of preference comparable to that of the reference signal.
\end{itemize}

The following contributions constitute the core of the present study. The first experiment explores the deterministic mapping between acoustic and semantic tokens. Then, two codecs have been selected for an analysis phase using a pretrained AnCoGen-Melspectrogram \cite{sadok2024ancogen} to determine how pitch, speaker identity, and linguistic content are encoded. Finally, two AnCoGen-Codec are trained as a plugin to enable direct prediction of speech attributes from tokens, facilitating both a deeper understanding and manipulation within the token space.


\begin{figure*}[h!]
  \centering
  \includegraphics[width=0.76\linewidth]{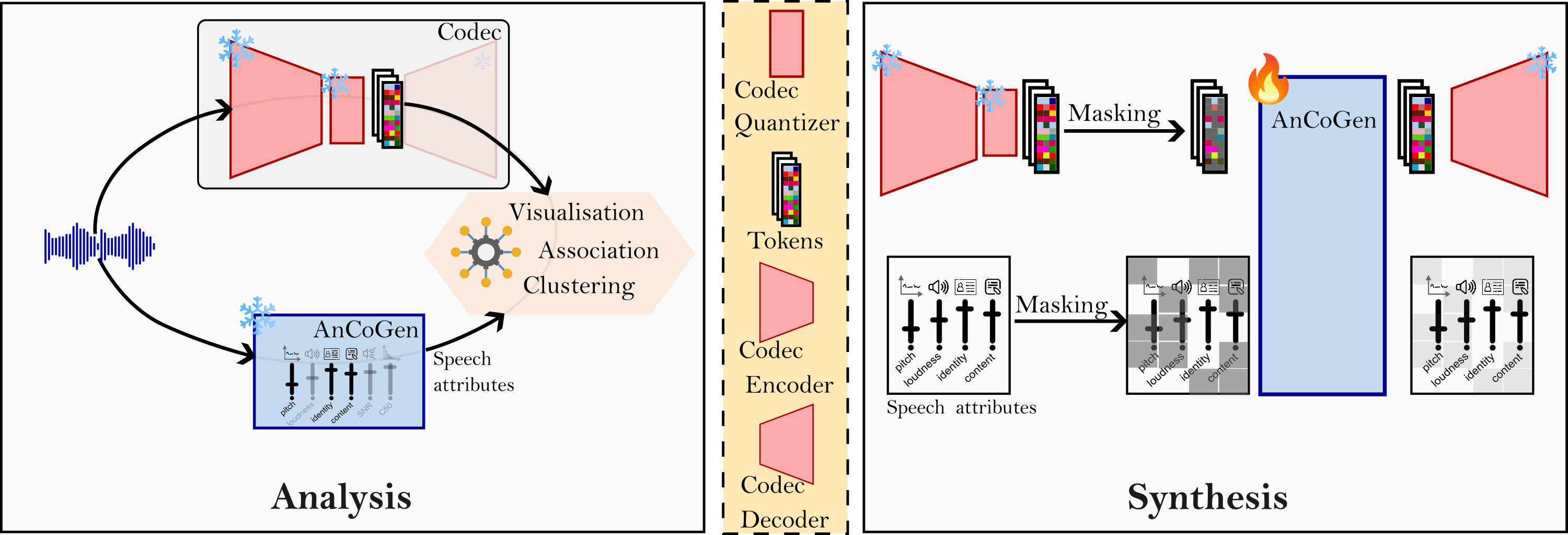}
  \caption{\textbf{Analysis (left):} Links between codec tokens and speech attributes are observed by performing two parallel forwards from the waveform. \textbf{Synthesis (right):} Once trained, AnCoGen model directly predicts  speech attributes from codec tokens and vice-versa}
  \label{fig:presentation}
\end{figure*}

\section{Analysis}

As illustrated in Figure~\ref{fig:presentation}, the analysis stage aims to delineate the coding of linguistic content, speaker identity, and pitch into neural codecs. To this end, we propose a framework in which AnCoGen-Melspectrogram \cite{sadok2024ancogen} predicts speech attributes (e.g., content and pitch) in parallel with codecs computing acoustic tokens. When the sampling rates of speech attributes and codecs differ, we eliminate some tokens from the longest sequence while maintaining time alignment to align the sequence length. In the following, the cardinality of the tokens of a given codec is denoted by $N_{\text{codec}}$. When a codec involves $M$ RVQ scales,  it consequently exhibits $\left(N_{\text{codec}}\right)^M$ possible code combinations for a given timestep, provided that the same cardinality is maintained across each RVQ scale.

\subsection{Content}
\label{sec:content}
\begin{figure}[t!]
    \centering
	\hfill
    \begin{subfigure}{0.236\textwidth}
        \centering
        \includegraphics[width=\textwidth]{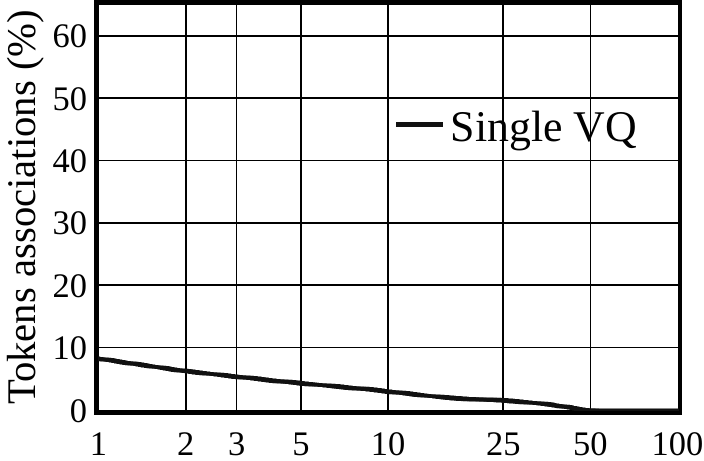}
        \caption{BigCodec}
        \label{fig:asso-bigcodec}
    \end{subfigure}
    \begin{subfigure}{0.2205\textwidth}
        \centering
        \includegraphics[width=\textwidth]{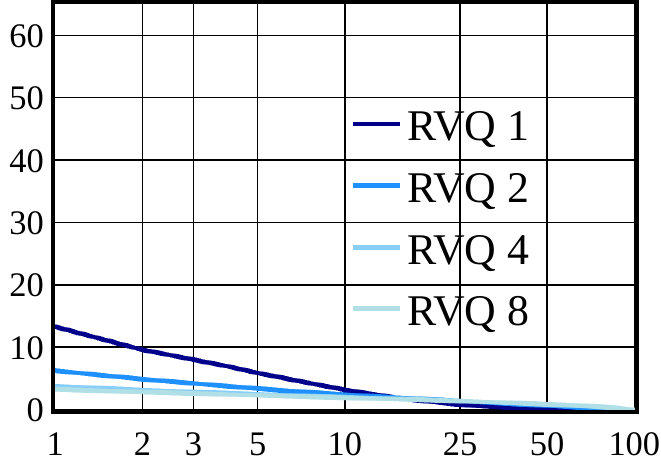}
        \caption{DAC}
        \label{fig:asso-dac}
    \end{subfigure}
	\hfill
    \begin{subfigure}{0.236\textwidth}
        \centering
        \includegraphics[width=\textwidth]{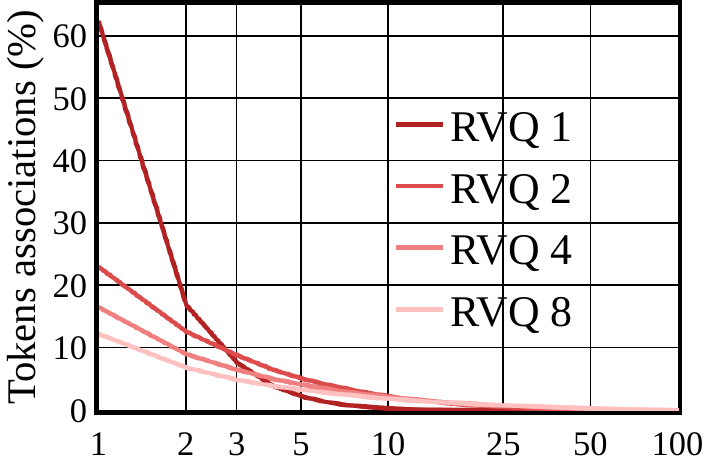}
        \caption{SpeechTokenizer}
        \label{fig:asso-speech_tok}
    \end{subfigure}
    \begin{subfigure}{0.2205\textwidth}
        \centering
        \includegraphics[width=\textwidth]{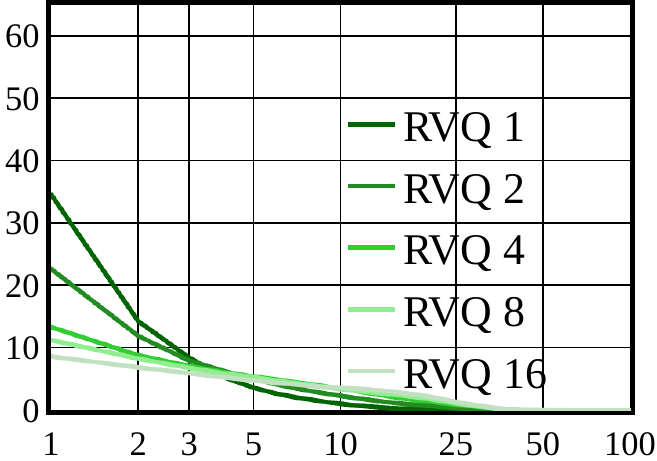}
        \caption{Mimi}
        \label{fig:asso-mimi}
    \end{subfigure}
    \caption{HuBERT/Codecs tokens associations on LibriSpeech test-clean : top-k HuBERT tokens usage per codec token/scale}
    \label{fig:asso}
\end{figure}

\noindent\textbf{Mapping between Hubert and Codec tokens:} A crucial preliminary inquiry pertains to the deterministic nature of the mapping of codec tokens on linguistic content, as represented by the $N_{\text{HuBERT}}=100$ tokens. For this purpose, the Librispeech clean test set \cite{panayotov2015librispeech} was tokenized, and the co-occurrences of acoustic tokens on every RVQ scale and Hubert tokens were observed. For each scale of each codec, a matrix of size $N_{\text{HuBERT}} \times N_{\text{codec}}$ is thus obtained. From this matrix, HuBERT tokens are ranked for each codec token based on their level of association, from highest to lowest. The results are then averaged for each codec and scale, expressed as percentages, and visualized in Figure~\ref{fig:asso}.

As might be anticipated, a purely deterministic mapping of token (\textit{i.e}., when a given code is systematically mapped to the same HuBERT token) is improbable due to the receptive fields of networks. Indeed, codes also gain meanings thanks to their surroundings, just as words in a sentence are not always translated to the same word in another language, depending on the context. Nevertheless, analyzing the basic one-to-one correspondence between acoustic and semantic tokens reveals several noteworthy properties. First, the single VQ scale of BigCodec exhibits a mean mapping of its token to its most associated HuBERT token of 8\%, with a long-tail distribution that results in a suboptimal mapping. This result can be justified by the fact that the codec employs a single scale and was not trained to align with HuBERT. A subsequent examination of DAC shows that the initial RVQ scale correlates more consistently with HuBERT. As scales increase, the mapping gets less precise, suggesting most linguistic content is encoded in the first scale. The first scale of SpeechTokenizer exhibits a high degree of alignment with HuBERT tokens, which is to be expected given that Zhang \emph{et al.} \cite{zhang2023speechtokenizer} distilled the HuBERT embeddings into this scale. All residual scales also display good associations, indicating they also contain linguistic content. At last, although Mimi introduced the splitRVQ tokenizer to restrict this linguistic leakage into higher RVQ scales, but the plot demonstrates that the larger RVQ scales still contain a considerable amount of linguistic content. This may be attributed to the extremely low frequency of Mimi, 12.5 Hz, which encompasses more HuBERT tokens.

\begin{figure*}[h!]
    \centering
    \begin{subfigure}[b]{0.176\textwidth}
        \centering
        \includegraphics[width=\textwidth]{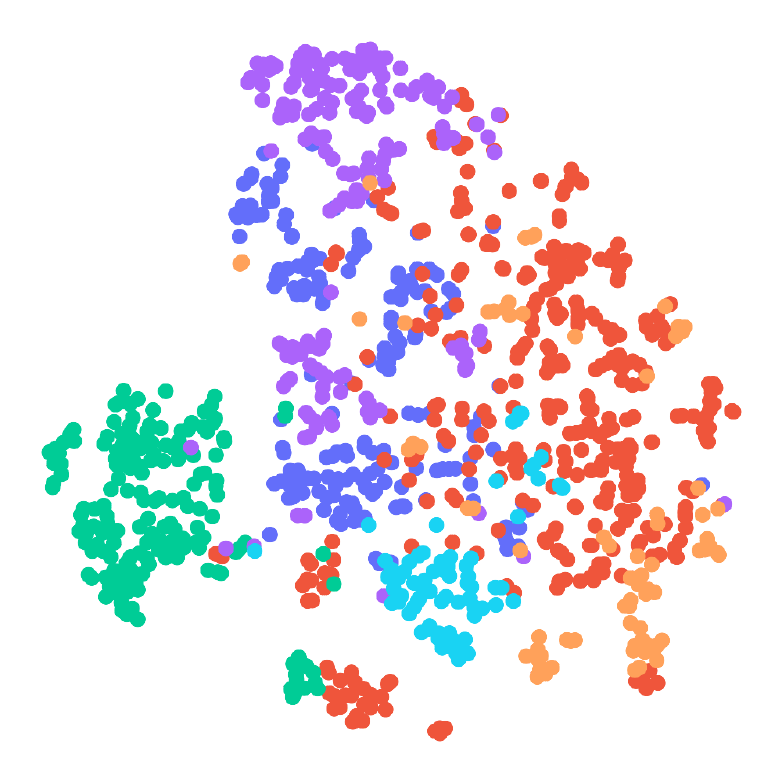}
        \caption{SpeechTokenizer RVQ1}
        \label{fig:content-speech-tok-rvq-1}
    \end{subfigure} \hfill
    \begin{subfigure}[b]{0.176\textwidth}
        \centering
        \includegraphics[width=\textwidth]{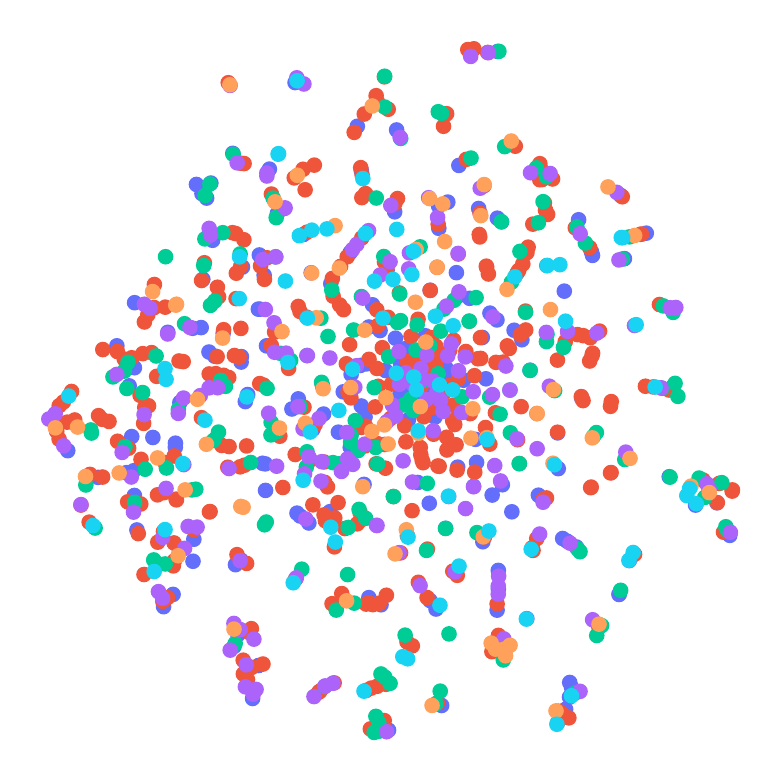}
        \caption{SpeechTokenizer RVQ2}
        \label{fig:content-speech-tok-rvq-2}
    \end{subfigure} \hfill
    \begin{subfigure}[b]{0.176\textwidth}
        \centering
        \includegraphics[width=\textwidth]{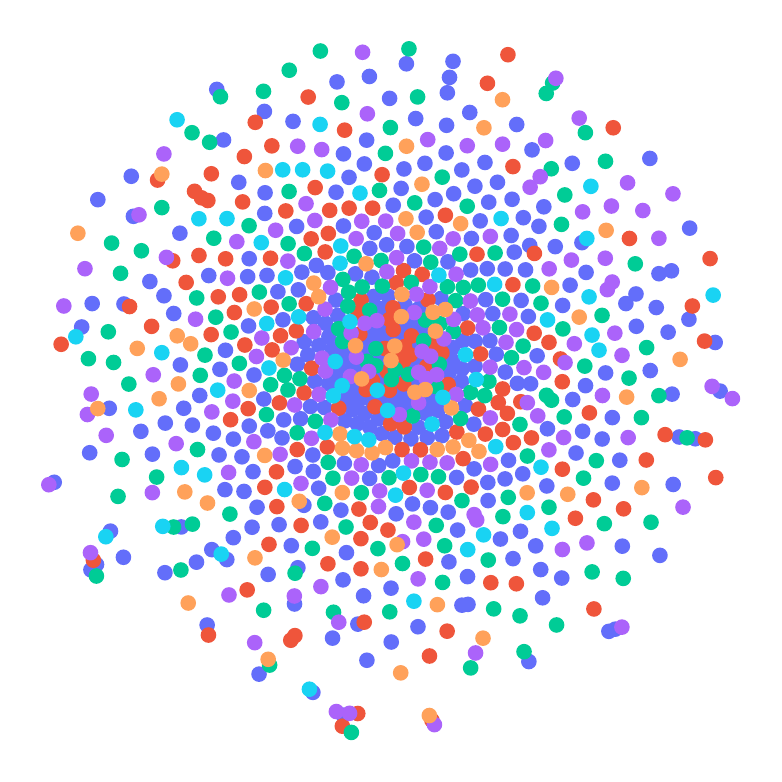}
        \caption{SpeechTokenizer RVQ8}
        \label{fig:content-speech-tok-rvq-8}
    \end{subfigure}
    \begin{subfigure}[b]{0.176\textwidth}
        \centering
        \includegraphics[width=\textwidth]{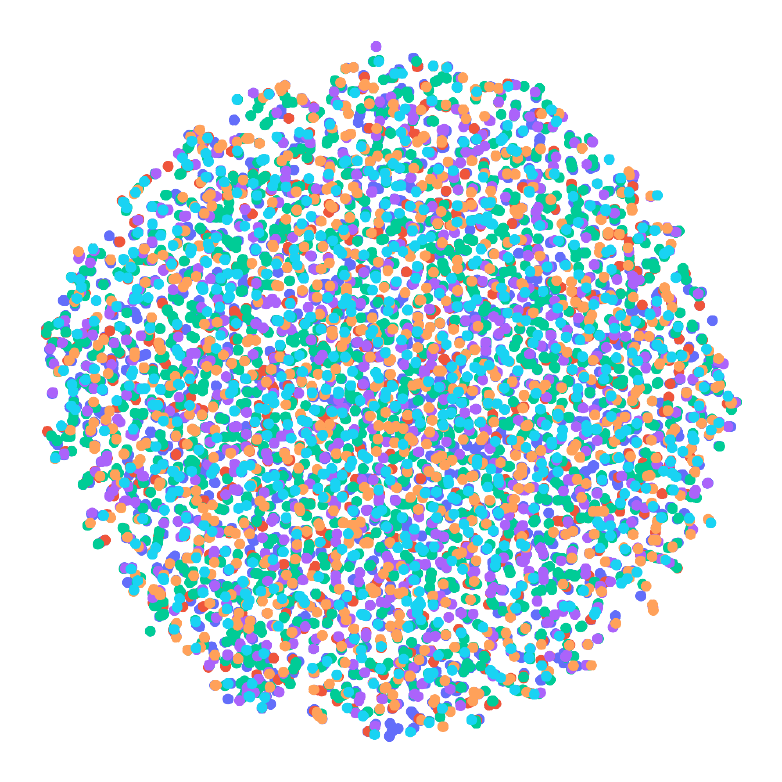}
        \caption{BigCodec}
        \label{fig:content-bigcodec}
    \end{subfigure} \hfill
        \includegraphics[width=0.1\textwidth]{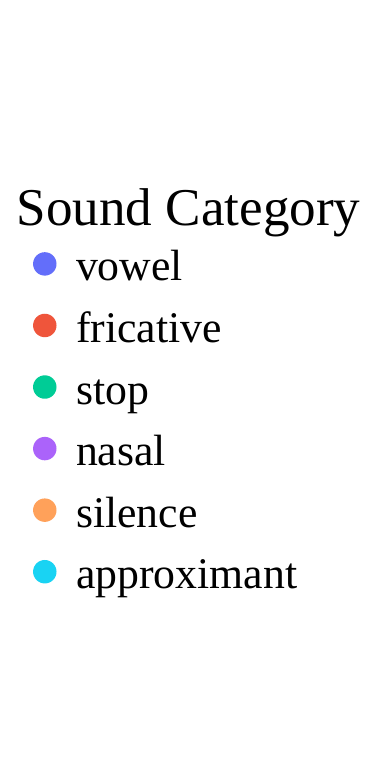}
 \hfill

    \caption{t-SNE visualisation. Color is attributed via codec-to-HuBERT and HuBERT-to-sound mappings.}
    \label{fig:content-HuBERT}
    \label{fig:TSNEcontent}
\end{figure*}

\noindent\textbf{Embeddings clustering:} in this analysis, our focus is narrowed to BigCodec and SpeechTokenizer, which present markedly contrasting association schemes. We posit a strong association between a specific codec token and its predominant HuBERT token. This assumption enables a two-component t-SNE \cite{van2008visualizing} visualization of learned token embeddings in the high-dimensional code space, as in Figure 3 of \cite{poli2024modeling}. The resulting t-SNE visualization is depicted in Figure~\ref{fig:TSNEcontent}, with the same perplexity applied to all graphs. The sound categories delineated in the dendogram of sound classification are presented in \cite{van2023rhythm}. This analysis enables us to assess the structural organization of the embedding space in relation to the linguistic content.

\begin{figure*}[h!]
    \centering
    \begin{subfigure}[b]{0.176\textwidth}
        \centering
        \includegraphics[width=\textwidth]{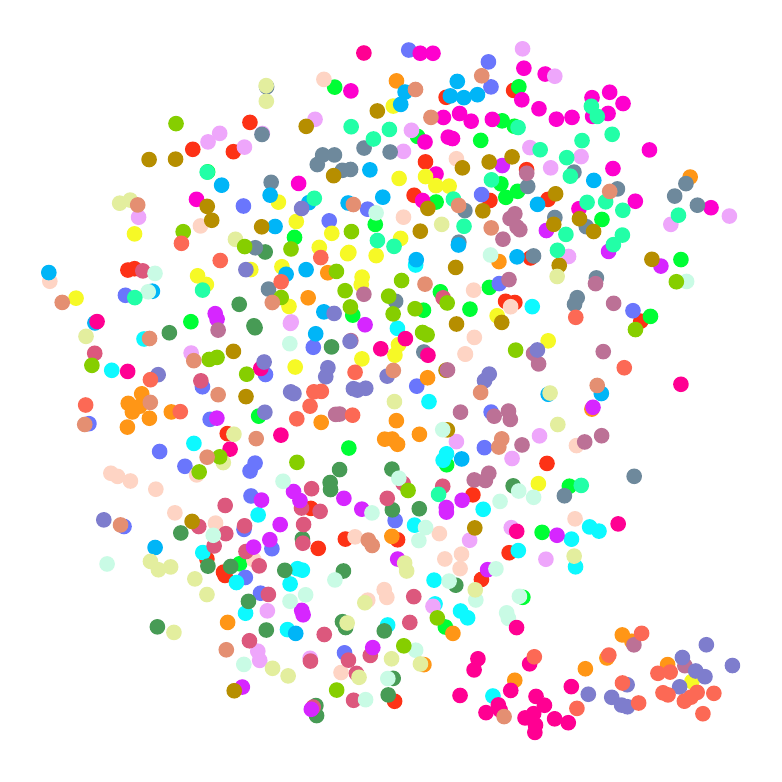}
        \caption{SpeechTokenizer RVQ1}
    \end{subfigure} \hfill
    \begin{subfigure}[b]{0.176\textwidth}
        \centering
        \includegraphics[width=\textwidth]{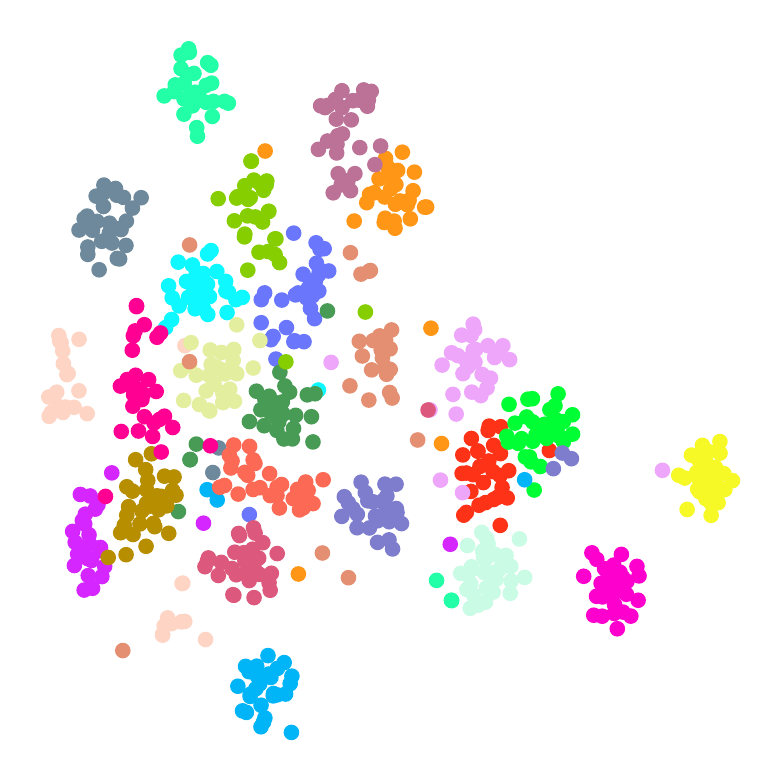}
        \caption{SpeechTokenizer RVQ4}
    \end{subfigure} \hfill
    \begin{subfigure}[b]{0.176\textwidth}
        \centering
        \includegraphics[width=\textwidth]{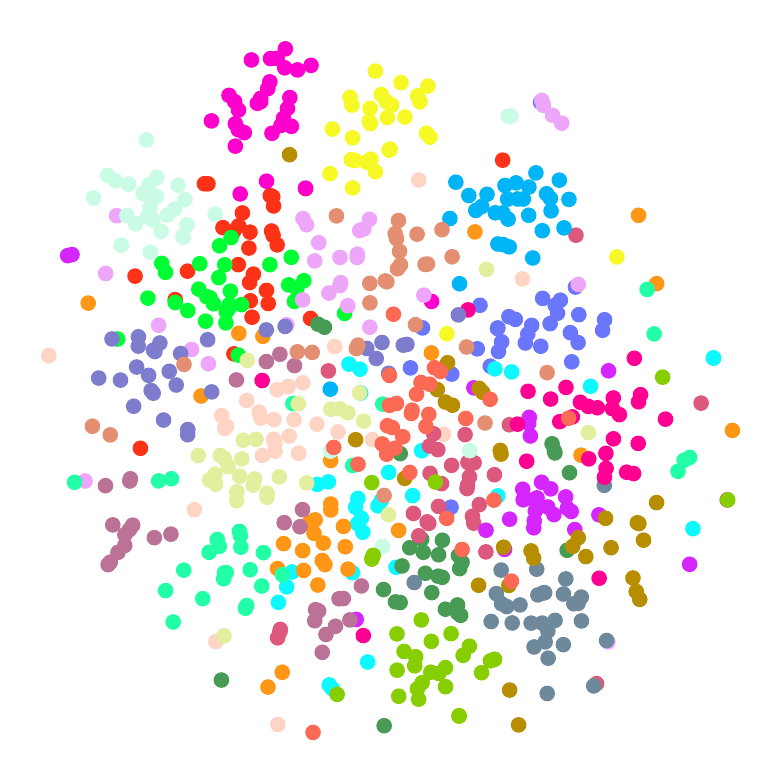}
        \caption{SpeechTokenizer RVQ8}
    \end{subfigure} \hfill
    \begin{subfigure}[b]{0.176\textwidth}
        \centering
        \includegraphics[width=\textwidth]{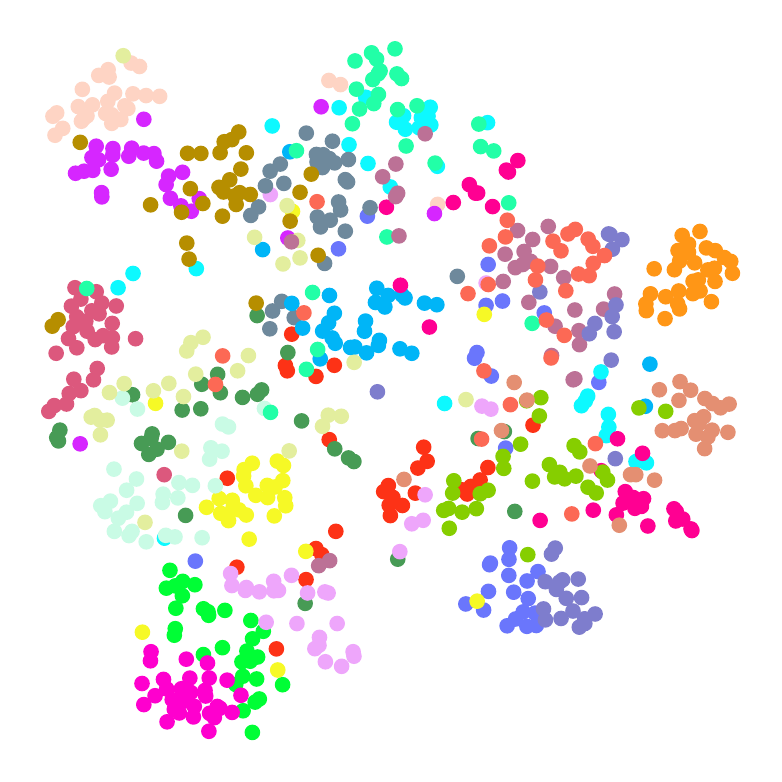}
        \caption{BigCodec}
    \end{subfigure}

    \caption{t-SNE visualisation of identity. Each dot corresponds to one utterance averaged over time, each color to one speaker.}
    \label{fig:TSNEidenity}
\end{figure*}

\begin{figure*}[h!]
    \centering
    \begin{subfigure}[b]{0.175\textwidth}
        \centering
        \includegraphics[width=\textwidth]{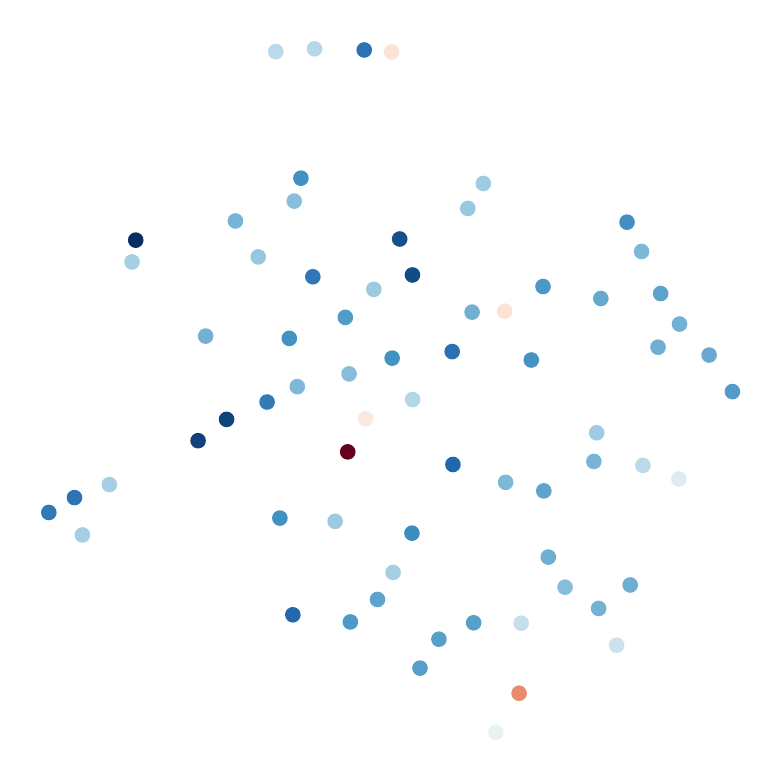}
        \caption{SpeechTokenizer RVQ1}
    \end{subfigure} \hfill
    \begin{subfigure}[b]{0.176\textwidth}
        \centering
        \includegraphics[width=\textwidth]{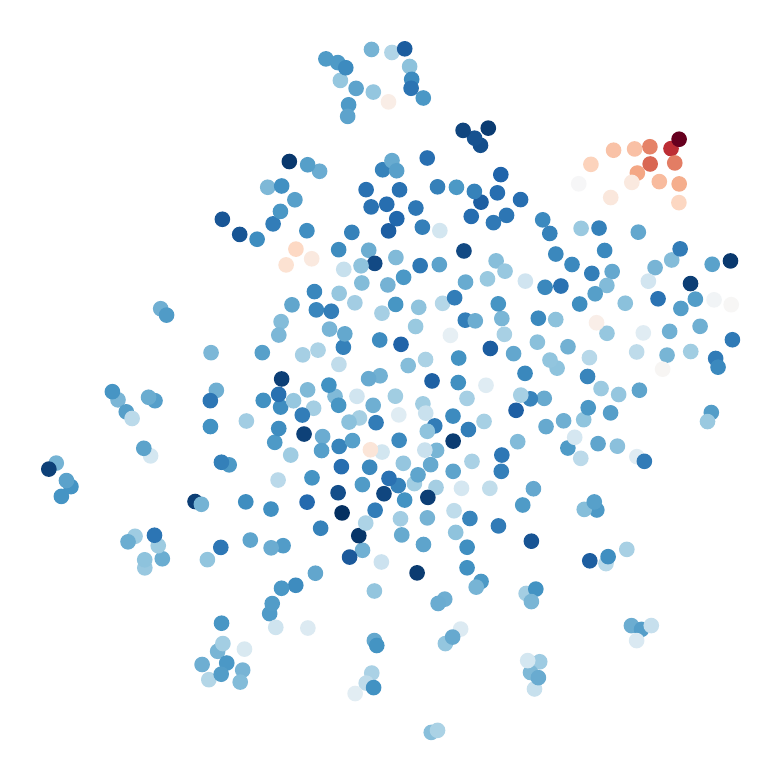}
        \caption{SpeechTokenizer RVQ2}
        \label{fig:mean_speech_tok_1_emb}
    \end{subfigure} \hfill
    \begin{subfigure}[b]{0.176\textwidth}
        \centering
        \includegraphics[width=\textwidth]{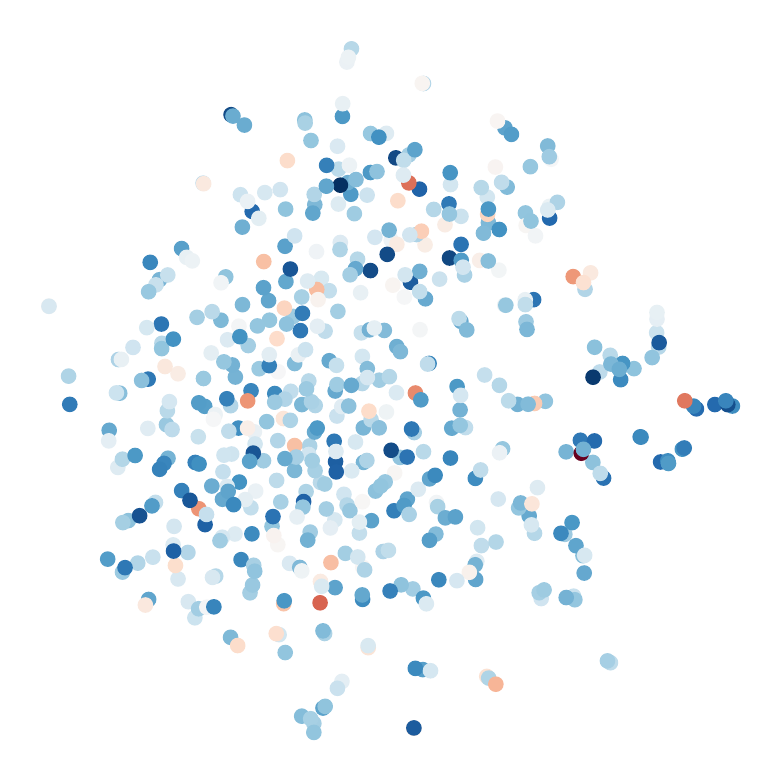}
        \caption{SpeechTokenizer RVQ8}
    \end{subfigure} \hfill
    \begin{subfigure}[b]{0.18\textwidth}
        \centering
        \includegraphics[width=\textwidth]{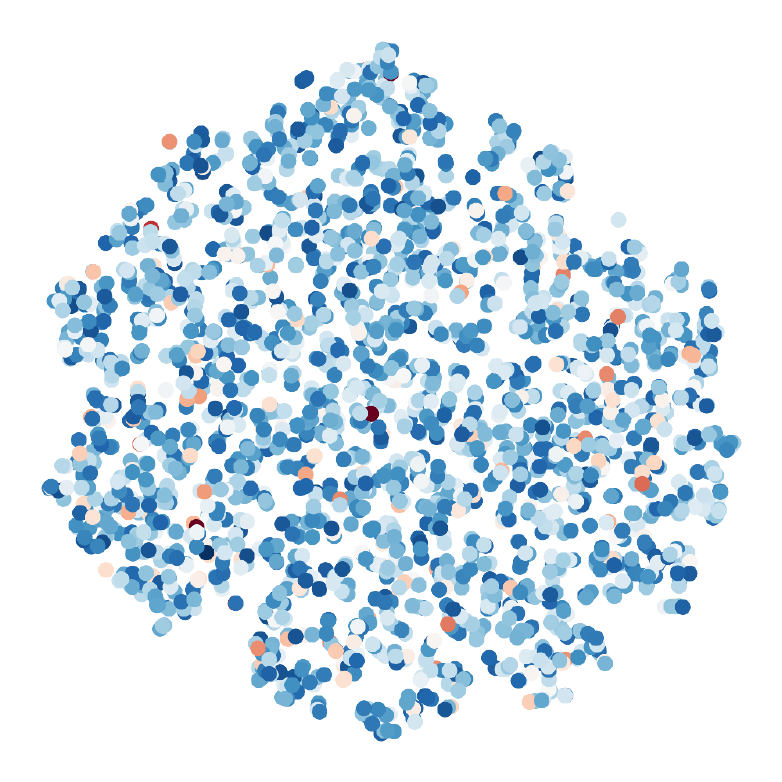}
        \caption{BigCodec}
    \end{subfigure}
        \includegraphics[width=0.06\textwidth]{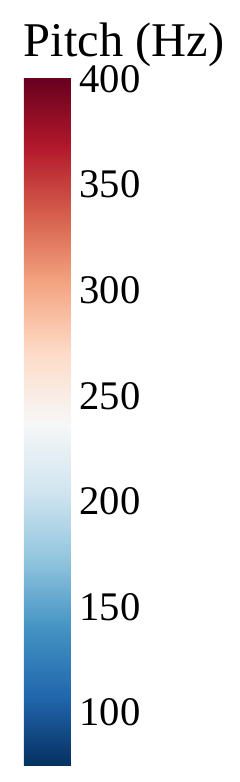}
  \hfill

    \caption{t-SNE visualisation of pitch.  Each dot corresponds to a token associated at least once with a specific (vowel) HuBERT token. The legend shows the mean pitch of all associations.}
    \label{fig:TSNEpitch}
\end{figure*}

As illustrated by the Figure~\ref{fig:content-speech-tok-rvq-1} plot, there is a clear delineation of clusters, indicative of the precise mapping between the initial scale of the SpeechTokenizer and HuBERT semantic tokens. In contrast, the Figures~\ref{fig:content-speech-tok-rvq-2} and \ref{fig:content-speech-tok-rvq-8} demonstrate an absence of this behavior.  The second RVQ scale exhibits some clusters, though they do not map to linguistic content. The eighth RVQ scale, on the other hand, appears much more diffuse, lacking any discernible structure. The BigCodec Figure \ref{fig:content-bigcodec}  exhibits a comparable behavior to the SpeechTokenizer second RVQ scale. It should be noted, however, that  $N_{\text{BigCodec}}>N_{\text{SpeechTokenizer}}$. This results in a greater number of clusters that do not align effectively with semantic content.

\subsection{Identity}

We also studied SpeechTokenizer and BigCodec with a two-component t-SNE visualization to examine how speaker identity is encoded. For this purpose, a subset of 720 utterances and 24 speakers from the LibriSpeech test clean corpus was tokenized, and the resulting embeddings were averaged on a per-utterance basis over time. A colormap depicting speaker identity was employed to obtain Figure \ref{fig:TSNEidenity}. Analysis of this figure reveals that speaker identity information is barely present in the first RVQ scale of SpeechTokenizer, is highly prevalent on the fourth scale, and is still present on the final scale, albeit with more dispersed clusters. For BigCodec, clustering is also observed, although some data points are outside the clusters. Finally, it is notable that clusters that are neighbors on one scale tend to maintain proximity on other scales or codecs.

\subsection{Pitch}

To analyze pitch structuration in the embedding space, the tokenized Librispeech test-clean dataset was filtered to retain only codec tokens associated with the 39th HuBERT token, which has high occurrence counts. This HuBERT token is identified as a vowel in \cite{van2023rhythm}. Pitch labels were computed as the mean pitch of these associations, with results shown in Figure~\ref{fig:TSNEpitch}. Pitch standard deviations are not displayed but are significant (approx. 40~Hz). Note that the number of points increases with the RVQ scale because the associations become less precise, as explained in Section~\ref{sec:content}. The analysis of Figure~\ref{fig:TSNEpitch} shows that the embeddings lack interpretability in terms of pitch, with the only visible feature being a small cluster of high pitch  values on the second RVQ scale of SpeechTokenizer (Figure~\ref{fig:mean_speech_tok_1_emb}). This highlights that pitch, being based on human auditory perception rather than the refinement of RVQ scales, is likely the most challenging speech attribute to decode from codec tokens.

\begin{figure}[b!]
  \centering
  \includegraphics[width=\linewidth]{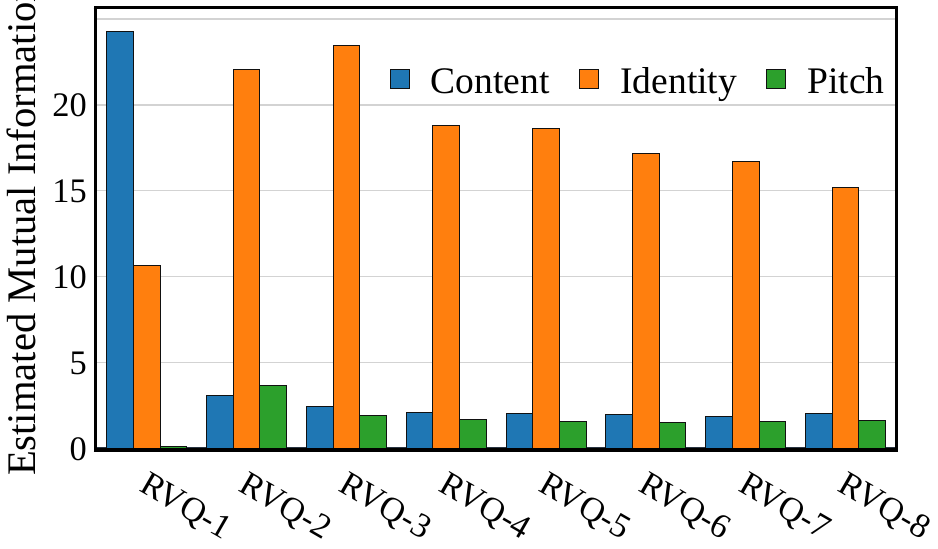}
  \caption{Mutual information between SpeechTokenizer RVQ scales and speech attributes}
  \label{fig:mi}
\end{figure}

\subsection{Mutual information}

A quantitative analysis was performed to evaluate the mutual information (MI) between SpeechTokenizer RVQ scales and speech attributes using the Contrastive Log-ratio Upper Bound (CLUB) method \cite{cheng2020club}. The resulting histogram (Figure~\ref{fig:mi}) displays non-normalized MI to highlight the relative scarcity of pitch MI compared to identity and content. For all attributes, the MI peaked where clusters were most distinct in the t-SNE visualizations, corroborating our previous observations.

\begin{figure*}[h!]
\centering
    \centering
    \resizebox{0.92\linewidth}{!}{
    \begin{tabular}{ccccccccccc}
    \toprule
         & & \multicolumn{2}{c}{\textbf{Analysis}} & \multicolumn{4}{c}{\textbf{Resynthesis}} & \multicolumn{2}{c}{\textbf{Control}}\\
        \cmidrule(lr){3-4} \cmidrule(lr){5-8} \cmidrule(lr){9-10}
	Task & & Pitch & Content &  \multicolumn{2}{c}{Non-intrusive} &\multicolumn{2}{c}{Intrusive} &  \multicolumn{2}{c}{Voice Conversion} & ~\\
	Metrics & Bitrate (kbps) & AAE \footnotesize{$\downarrow$} & Acc. \footnotesize{$\uparrow$} & N-MOS \footnotesize{$\uparrow$} & BAK \footnotesize{$\uparrow$} & STOI \footnotesize{$\uparrow$} & BertScore \footnotesize{$\uparrow$} & N-MOS \footnotesize{$\uparrow$} & COS \footnotesize{$\uparrow$} & ~ \\
	 \midrule
\vspace{0.05cm}
	AnCoGen - \emph{Melspectrogram}  \cite{sadok2024ancogen}& 51.2 & \textbf{4.60} & \underline{81.20} & 4.26 &  \textbf{4.16} & \textbf{0.79} & \underline{0.87} & 4.24& \underline{0.72} \\\vspace{0.05cm}
	AnCoGen - \emph{BigCodec}  & 1.04 & 7.04 & 73.58 & \textbf{4.39} & \underline{4.07} & 0.72 & 0.82 & \textbf{4.37} & 0.68 \\
	AnCoGen - \emph{SpeechTokenizer}  & 4.00 & \underline{6.48} & \textbf{82.00} & \underline{4.34} & 3.91 & \underline{0.75} & \textbf{0.88} & \underline{4.26} & \textbf{ 0.74} \\
        \bottomrule
    \end{tabular}
    }
    \captionof{table}{AnCoGen results (best score in each column is in bold, second best score is underlined).}
    \label{tab:tasks}
\end{figure*}

\section{Synthesis}

The previous section highlighted that key speech features—pitch, identity, and linguistic content—are entangled in the codec representation, limiting the interpretability of encoded speech signals. Only a few codecs stand out as exceptions to this trend \cite{bie2024learning,zheng2024freecodec,ju2024naturalspeech}. Inspired by \cite{polyak2021speech, gengembre2024disentangling, sadok2024ancogen}, we introduce a bidirectional framework to analyze, control, and generate speech audio from codec tokens. This approach maps codec tokens to speech attributes (analysis) and reverses the process to map attributes back to codec tokens (synthesis). As in the previous section, our study focuses on \emph{SpeechTokenizer} and \emph{BigCodec}, chosen for their contrasting design paradigms (see Section~\ref{sec:intro}). All AnCoGen-Codec models are trained on the LibriSpeech-100-clean dataset, with performance evaluations conducted primarily on the LibriSpeech test dataset.

\vspace*{-.2cm}
\subsection{Method}

Our AnCoGen-Codec utilizes two speech signal representations: codec tokens and four high-level attributes capturing linguistic, prosodic, and acoustic features. These attributes include \emph{linguistic content} (from HuBERT encoder outputs), \emph{pitch contour} (extracted using CREPE \cite{kim2018crepe}), \emph{loudness} (computed via root mean square signal), and \emph{speaker identity} (derived from embeddings of the pre-trained ECAPA-TDNN model \cite{desplanques2020ecapa}). For further details, refer to the original AnCoGen paper \cite{sadok2024ancogen}.

\emph{During training}, the input speech signal is converted into codec and speech attribute token sequences, which are partially masked following a predefined strategy. AnCoGen, leveraging an encoder-decoder Transformer, embeds visible tokens into learned vectors processed by the encoder using multi-head self-attention. Mask tokens are added, and the sequence is fed to the decoder which predicts masked token indices thanks to several linear layers for multi-scale codecs. Training optimizes the cross-entropy loss between predicted and ground-truth indices. \emph{At inference}, AnCoGen supports both speech analysis and generation. For analysis, the codec token sequence of an input speech signal is fed to the encoder with all speech attribute tokens masked, allowing the decoder to predict the corresponding speech attributes. For generation, the encoder processes the speech attribute token sequence while all codec tokens are masked. The decoder predicts the codec tokens, which are then used to reconstruct the audio signal. Finally, the frozen codec decoder converts the reconstructed codec tokens into an audio waveform.

\vspace*{-.2cm}
\subsection{Results}
This section discusses the results presented in Table~\ref{tab:tasks} for three tasks: pitch and content estimation (analysis task), resynthesis task, and voice conversion (control task).

\noindent\textbf{Speech analysis:} This experiment evaluates AnCoGen's ability to estimate speech attributes, such as pitch and content, from codec token representations. Pitch estimation is evaluated on the PTDB dataset \cite{pirker2011pitch} with ground-truth pitch values, using the Average Absolute Error (AAE) metric. While content estimation is performed on the LibriSpeech test set, measuring accuracy across 100 classes defined by HuBERT. Both BigCodec and SpeechTokenizer tokens in AnCoGen yield strong pitch estimation results, with SpeechTokenizer showing slightly better accuracy on average. For content analysis, AnCoGen achieves its highest performance with SpeechTokenizer tokens, reaching 82$\%$ accuracy. This outcome aligns with expectations, as SpeechTokenizer incorporates HuBERT's semantic information to regularize the first RVQ scale.

\noindent\textbf{Speech resynthesis:} In this experiment, we use AnCoGen to map back and forth between codec tokens and speech attributes, evaluating potential information loss. We use two non-intrusive audio quality metrics: DNSMOS Background Noise Quality (BAK) \cite{reddy2022dnsmos} and Noresqa MOS (N-MOS) \cite{manocha2021noresqa, kumar2023torchaudio}, along with two intrusive metrics: Short-Time Objective Intelligibility (STOI) \cite{taal2010short} and speechBertScore \cite{saeki2024speechbertscore}. As a reference, we compare against the codec decoder output without AnCoGen to isolate potential artifacts. The three AnCoGen models—Melspectrogram, BigCodec, and SpeechTokenizer—show good synthesis quality across the four speech attributes. AnCoGen-BigCodec achieves the highest N-MOS score (4.39) but demonstrates less accurate content reconstruction due to BigCodec's content analysis errors. AnCoGen-SpeechTokenizer offers good listening quality but introduces frame artifacts, as seen in BAK scores, due to parallel pattern scale prediction \cite{copet2024simple}. However, SpeechTokenizer excels in content reconstruction, reflecting the strong linguistic interpretability of its first RVQ scale.

\noindent\textbf{Speech control:} In this task, we investigate voice conversion using AnCoGen through the \emph{analysis-control-generation} framework. During the control step, the speaker identity of the source signal is replaced with that of the target while preserving all other source attributes. Speech signals from 10 randomly selected identities in the LibriSpeech test set are used for evaluation. Speaker similarity is measured using cosine similarity (COS) computed with Resemblyzer embeddings~\cite{wan2018generalized}, while audio quality is assessed using the N-MOS metric. All three approaches achieve good voice conversion quality, with AnCoGen-SpeechTokenizer delivering the highest target voice similarity, evidenced by a COS score of 0.74.

\section{Conclusion}
This work studies two prominent neural audio codecs, assessing their ability to represent content, pitch, and speaker identity. Our visual and quantitative analysis reveals that these speech attributes are entangled within the codecs' quantized latent spaces, limiting interpretability - even for codecs designed with disentanglement in mind. We also introduce a framework inspired by AnCoGen, which bridges these codecs to improve interpretability. This bidirectional approach links codec tokens to key speech attributes (e.g., pitch, content, loudness, identity), enabling their extraction for analysis and speech generation.

\section{Acknowledgements}

This work was granted access to the HPC/AI resources of [CINES / IDRIS / TGCC] under the grant 2022-AD011013469 awarded by GENCI and partially funded by the French National Research Agency under the ANR Grant No. ANR-20-THIA-0002.

\bibliographystyle{IEEEtran}
\bibliography{mybib}

\end{document}